\documentclass[12pt]{iopart}
\usepackage[utf8]{inputenc}
\usepackage{graphicx}% Include figure files
\usepackage{float}
\usepackage{caption}
\usepackage{subcaption}
\usepackage{xcolor}
\usepackage{dcolumn}% Align table columns on decimal point
\usepackage{bm}% bold math
\usepackage{tabularx}
\usepackage{booktabs}
\usepackage[colorlinks=true,linkcolor=blue,urlcolor=blue,citecolor=blue]{hyperref}
\usepackage{multirow}
\usepackage{siunitx} 
\usepackage{iopams}
\usepackage{algorithm}
\usepackage{algpseudocode}

\newcommand{\bfrt}{\bm{r},t}

\newcommand{\ubr}{\underline{\bm{r}}}

\newcommand{\s}{_\mathrm{{\scriptscriptstyle S}}}
\newcommand{\h}{_\mathrm{{\scriptscriptstyle H}}}
\newcommand{\xc}{_\mathrm{{\scriptscriptstyle XC}}}

\newcommand{\ext}{_\mathrm{{\scriptscriptstyle ext}}}

\newcommand{\Eqref}[1]{Eq.~(\ref{#1})}

\DeclareSIUnit{\au}{a.u.}

\begin{document}

\title{Machine Learning Time Propagators for Time-Dependent Density Functional Theory Simulations}

\author{Karan Shah$^{1,2}$, Attila Cangi$^{1,2,*}$,}
\address{$^1$ Center for Advanced Systems Understanding, 02826 G\"orlitz, Germany}
\address{$^2$ Helmholtz-Zentrum Dresden-Rossendorf, 01328 Dresden, Germany}
\address{$*$  a.cangi@hzdr.de}

\vspace{10pt}
\begin{indented}
\item[]\today
\end{indented}

\begin{abstract}
    Time-dependent density functional theory (TDDFT) is a widely used method to investigate electron dynamics under external time-dependent perturbations such as laser fields. In this work, we present a machine learning approach to accelerate electron dynamics simulations based on real time TDDFT using autoregressive neural operators as time-propagators for the electron density. By leveraging physics-informed constraints and featurization, and high-resolution training data, our model achieves superior accuracy and computational speed compared to traditional numerical solvers. We demonstrate the effectiveness of our model on a class of one-dimensional diatomic molecules under the influence of a range of laser parameters. This method has potential in enabling on-the-fly modeling of laser-irradiated molecules and materials by utilizing fast machine learning predictions in a large space of varying experimental parameters of the laser.
\end{abstract}

\section{Introduction}
\label{sec:intro}

Time-Dependent Density Functional Theory (TDDFT) \cite{runge_density-functional_1984} is a widely used method to study the evolution of electronic structure under the influence of external time-dependent potentials. It is used to calculate various excited state properties such as excitation energies \cite{adamo_calculations_2013}, charge transfer \cite{maitra_charge_2017}, stopping power \cite{yost_examining_2017}, optical absorption spectra \cite{jacquemin_excited-state_2011} and non-linear optical properties \cite{goncharov_non-linear_2014}. Due to its favorable balance between accuracy and computational cost, TDDFT has been applied in various domains including photocatalysis \cite{herringRecentAdvancesRealTime2023}, biochemistry \cite{varsanoTDDFTStudyExcited2006}, nanoscale devices \cite{senanayakeRealTimeTDDFTInvestigation2019} and the study of light-matter interactions in general \cite{neufeldBenchmarkingFunctionalsStrongField2024}.

For weak perturbations, the linear response formalism of TDDFT is used to calculate the excitation spectrum of a system ~\cite{PGB00}. It is calculated using the Casida equation \cite{casida_time-dependent_1995} as an eigenvalue problem that describes the first order response of the density. In contrast, the electron density is directly propagated in time under the real time formalism. Real time TDDFT can be used to calculate the nonlinear response of the density under strong perturbations such as ultrafast electron dynamics with strong laser fields \cite{provorse_electron_2016}.

There are multiple components that must be decided for real time TDDFT calculations. These include preparation of the initial density and orbitals, choice of the exchange-correlation functional, the form of the external time-dependent potential which is determined by the problem to be solved and the choice of the time-propagation scheme. The time evolution is a significant fraction of the computational cost \cite{castro_propagators_2004, gomez_pueyo_propagators_2018}.

Machine learning (ML) for accelerating scientific simulations is a rapidly growing area of research \cite{carleo_machine_2019}. A variety of unsupervised and data-driven models have been developed to solve differential equations across a wide range of domains \cite{karniadakisPhysicsinformedMachineLearning2021a}. 

Neural operators (NOs) \cite{kovachki_neural_2023}
are a class of models that map function-to-function spaces, as opposed to the finite-dimensional vector space mappings of neural networks. This is especially useful for partial differential equation (PDE) problems where experimental or simulation data is available. Fourier neural operators (FNOs) \cite{liFourierNeuralOperator2020} are a type of NO which represent operator weights in Fourier space. The main advantages of FNOs are that they generalize well across function spaces, and being resolution-invariant, they can be used for inference on higher resolution grids than the training set grids. FNOs have also been used for forward and inverse PDE problems in various domains \cite{azizzadenesheli_neural_2024}.

While many ML models have been developed for the electronic structure in ground state DFT \cite{fiedler_deep_2022,brockherde_bypassing_2017,eickenberg_solid_2018,schmidt_learning_2018,MiRy2019,grisafi_transferable_2019, fabrizio_electron_2019,chandrasekaran_solving_2019, tsubaki_quantum_2020,ben_mahmoud_learning_2020,lewis_learning_2021,ellis_accelerating_2021,FiMoScVo23,Rackers_2023,Shao_2023,CaFiBrSh25},
there have been relatively fewer efforts focused on TDDFT. Some applications of ML for TDDFT include development of exchange-correlation potentials \cite{ScBeMa19,suzuki_machine_2020,yang_machine-learning_2023,BhGuIs24} and predicting properties such as spectra~\cite{BoShZhXu24,ShLe25} and stopping power \cite{WaBlLeMa24}. In addition, there is prior work on developing ML models for quantum dissipative systems~\cite{Zh25,zhangNeuralQuantumPropagators2025,zhangNeuralNetworkSolution2025} and for FNO based TDDFT time propagators for model systems under the influence of a fixed laser field~\cite{shahAcceleratingElectronDynamics2024}.

In this work, we demonstrate the effectiveness of FNOs in propagating the electron density in time under the TDDFT framework. Instead of propagating orbitals in time as done conventionally in terms of the time-dependent Kohn-Sham equations, we use the FNO propagator to directly evolve the density. This has two advantages: the computational cost does not scale with the number of orbitals and larger propagation time steps can be used, thus using fewer iterations. Section~\ref{sec:methods} contains a brief description of TDDFT, FNOs and our proposed autoregressive model. Section~\ref{sec:results} contains results obtained with this ML model for the time evolution of one-dimensional diatomic molecules under the influence of a laser pulse in the dipole approximation. We show that the model can be generalized across ionic configurations, is faster and more accurate than comparable numerical simulations and can also be used for higher resolution grids. We discuss the physical viability of the predicted densities in Section~\ref{sec:discussion}, and then conclude in Section~\ref{sec:conclusion}. Additional details simulation and machine learning parameters are provided in \hyperref[sec:SI]{SI}.
We use atomic units a.u. ($\hbar = m_e = e = 1)$ unless specified otherwise.

\section{Methods}
\label{sec:methods}

\subsection{Time-Dependent Density Functional Theory}
The time-dependent many-body Schr\"odinger equation describes the dynamics of systems composed of multiple interacting particles. A common problem is interacting electrons in atoms and molecules, for which the Schr\"odinger equation is
\begin{equation}
i \frac{\partial}{\partial t} \Psi(\ubr, t)=\hat{H}(t) \Psi(\ubr, t),
\label{eq:tdse}
\end{equation}
where $\ubr$ represents the collective coordinates of the $N$ electrons, $\Psi(\ubr, t)$ is the many-body wavefunction and $\hat{H}$ is the Hamiltonian operator, representing the total energy of the system. For this problem the form of $\hat{H}$ is 
\begin{equation}
\hat{H}(t)=\hat{T}+\hat{W}+\hat{V}(t),
\label{eq:tdham}
\end{equation}
where $\hat{T}=\sum_{j=1}^N-\nabla_j^2/2$ is the kinetic energy operator, $\hat{W}=\sum_{1\le j<k\le N} \frac{1}{\left|\mathbf{r}_j-\mathbf{r}_k\right|}$ represents the electron-electron interaction, and $\hat{V}(t)=v\ext(\bfrt)$ is the time-dependent external potential operator which includes the static external potential due to the ions and an external time-dependent potential that drives the system. Note that in writing this Schr\"odinger equation we are assuming the Born-Oppenheimer approximation, i.e. we are treating the dynamics of the ions as occurring on a much longer time scale than the dynamics of the electrons, and can therefore treat the ions as classical point-like particles. The initial state $\Psi(\ubr) = \Psi(\ubr, t_0)$ is obtained by solving the time-independent Schr\"odinger equation
\begin{equation}
    \hat{H}\Psi(\ubr) = E \Psi(\ubr)
    \label{eq:tise}
\end{equation}
which is an eigenvalue problem where $E$ denotes the eigenvalues and corresponds to the total energy of the system. The static Hamiltonian $\hat{H}$ is similar in form to \Eqref{eq:tdham} but with a time-independent external potential $\hat{V}$.

The many-body wavefunction $\Psi$ contains all the information about the system and can be used to calculate its properties. However for 3 spatial dimensions, we need to solve a system of $3N$ variables and the computational cost scales exponentially with $N$, rendering it intractable for all but the simplest systems. 

DFT and TDDFT allow us to solve \Eqref{eq:tise} and \Eqref{eq:tdse} respectively by reformulating the problem in terms of electron density rather than the wavefunction, dramatically reducing the computational cost.

TDDFT is based on the Runge-Gross theorem \cite{runge_density-functional_1984} which states that for a system with a given ground state many-body wavefunction $\Psi_0=\Psi_0\left(\boldsymbol{r}, t_0\right)$, there exists a unique mapping between the potential and the time-dependent density. The density of the interacting many-body system can be obtained by solving a system of fictitious non-interacting particles governed by the time-dependent Kohn-Sham (TDKS) equations \cite{van_leeuwen_causality_1998}
\begin{equation}
\hat{H}_{\text{KS}}\,\phi_j(\bm{r},t) \;=\; i \frac{\partial \phi_j(\bm{r},t)}{\partial t}, \quad j = 1,\ldots, N,
\label{eq:kseqntd}
\end{equation}
where the electron density 
\begin{equation}
n(\bm{r}, t) = \sum_{j=1}^N |\phi_j(\bm{r}, t)|^2
\label{eq:ksdens}
\end{equation}
is the quantity of interest. The Kohn-Sham Hamiltonian is defined as
\begin{equation}
\hat{H}_{\text{KS}} = -\frac{1}{2}\nabla^2 + v_s[n](\bm{r},t),
\label{eq:ksham}
\end{equation}
where the Kohn-Sham potential $v_s[n](\bm{r},t)$ is a functional of the density $n(\bm{r},t)$. The correspondence between density and potential is established through
\begin{equation}
v\s[n](\bfrt) = v\ext(\bfrt) + v\h(n(\bfrt)) + v\xc[n](\bfrt),
\label{eq:kspot}
\end{equation}
with $v\ext(\bfrt)$ the external potential, $v\h(\bfrt)$ the Hartree potential, and $v\xc(\bfrt)$ the exchange-correlation potential. The external potential is composed of the ionic potential and any time-dependent external perturbation such as a laser field.

While in theory, $v\xc(\bfrt)$ depends on the density at all previous time steps, the adiabatic approximation is often used in practice \cite{ullrich_time-dependent_2012}. We use the adiabatic local density approximation (ALDA).
Under ALDA, \( v\xc(\bfrt) \) is approximated as:
\begin{equation}
 v\xc^{\text{ALDA}}[n, \Psi_0, \Phi_0](\mathbf{r}, t) =  v\xc^{\text{LDA}}(n(\mathbf{r}, t)),
 \label{eq:ksalda}
\end{equation}
where \( v\xc^{\text{LDA}} \) is the LDA exchange-correlation potential in ground-state DFT. 

These coupled equations \Eqref{eq:kseqntd} to \Eqref{eq:kspot} are solved through iterative numerical algorithms \cite{castro_propagators_2004}.

\subsubsection{Time Propagators} \label{tddft_tp}
The general form of the time evolution operator $\hat{U}$ for a time domain $T $ is given by:
\begin{equation}
\phi_i(\mathbf{r}, T) = \hat{U}(T, t_0) \phi_i(\mathbf{r}, t_0).
\end{equation}
When the Hamiltonian $\hat{H}_{KS}$ is time-independent, $\hat{U}(T, t_0) = \exp(-i \hat{H}_{KS} (T-t_0))$. However, in the general case, when external fields are present or due to the inherent time-dependence of the density within the Hamiltonian functional, $\hat{H}_{KS}(t)$ is time-dependent. The solution then requires a time-ordered exponential:
\begin{equation}
\hat{U}(T, t_0) = \mathcal{T} \exp\left(-i \int_{t_0}^{T} \hat{H}_{KS}(t') dt'\right),
\end{equation}
where $\mathcal{T}$ is the time-ordering operator.

Evaluating the time-ordered exponential directly is generally intractable. In practice, the total time evolution is broken down into a sequence of short time steps $\Delta t$. The total evolution operator over a time domain $T$ starting from $t_0=0$ is approximated as a product of short-time propagators:
\begin{equation}
\hat{U}(T, t_0) = \prod_{j=t_0}^{N-1} \hat{U}(t_{j+1}, t_j),
\end{equation}
with $t_0=0$, $t_{j+1}=t_j+\Delta t_j$, and $t_N=T$. Usually, a constant time step $\Delta t$ is used. The accuracy and stability of the simulation depend crucially on the approximation used for the short-time propagator $\hat{U}(t_j+\Delta t, t_j)$.

A fundamental requirement for the exact evolution operator is unitarity: $\hat{U}^{\dagger}(t+\Delta t, t) = \hat{U}^{-1}(t+\Delta t, t)$. This property ensures that the norm of the wavefunctions is conserved over time, $\langle\phi_i(t)|\phi_i(t)\rangle = \text{const}$, which in turn guarantees the conservation of the total number of electrons (i.e., the integrated density $\int n(\mathbf{r}, t) d\mathbf{r}$). Another important property is time-reversal symmetry: $\hat{U}(t+\Delta t, t) = \hat{U}^{-1}(t, t+\Delta t)$, meaning that propagating forward by $\Delta t$ is the inverse of propagating backward by $\Delta t$. A robust numerical time propagation algorithm should ideally preserve these properties, or approximate them sufficiently well to ensure physical conservation laws and stability over the desired simulation time.

Numerous methods have been developed and analyzed for approximating the short-time propagator $\hat{U}(t_j+\Delta t, t_j)$ \cite{castro_propagators_2004, gomez_pueyo_propagators_2018}, including Magnus expansions that approximate the exponent in $\hat{U}(\Delta t)=\exp(\hat{\Omega}(\Delta t))$, split-operator schemes effective when the Hamiltonian separates into analytically solvable parts (e.g., kinetic and potential), explicit and implicit Runge–Kutta methods adapted to the Schr\"odinger equation, exponential-midpoint and other exponential integrators, and implicit schemes such as Crank–Nicolson \cite{Crank_Nicolson_1947}.
The choice of propagator often involves a trade-off between accuracy, stability (especially concerning the maximum usable time step $\Delta t$), computational cost per step, and conservation properties.

For this work, the reference data for training our neural network model was generated using the Crank-Nicolson method \cite{Crank_Nicolson_1947}. This is an implicit method, formulated as:
\begin{eqnarray}
\left( 1 + \frac{i \Delta t}{2} \hat{H}_{KS}(t+\Delta t/2) \right) \phi_i(t+\Delta t) 
\approx {} & \nonumber \\
\left( 1 - \frac{i \Delta t}{2} \hat{H}_{KS}(t+\Delta t/2) \right) \phi_i(t)
\end{eqnarray}
where $\hat{H}_{KS}(t+\Delta t/2)$ is typically an approximation of the Hamiltonian at the midpoint of the time step. The Crank-Nicolson propagator, $\hat{U}_{CN}(\Delta t) \approx [1 + i \hat{H}_{KS} \Delta t / 2]^{-1} [1 - i \hat{H}_{KS} \Delta t / 2]$, is unitary and stable, although its implicit nature requires solving a linear system at each time step.

While these traditional methods propagate the Kohn-Sham orbitals $\phi_i$, our approach deviates from this standard procedure. We train a Fourier Neural Operator (FNO) to directly propagate the electron density $n(\mathbf{r}, t)$ in time with larger $\Delta t$, bypassing the need for explicit time-propagation of the Kohn-Sham orbitals.

\subsubsection{Model Systems}

We consider an external potential 
\begin{equation}
    v_{\text{ext}}(\mathbf{r}, t) = v_{\text{ion}}(\mathbf{r}) + v_{\text{las}}(t)\,,
\end{equation}
where we simulate a class of one-dimensional diatomic molecules that serves as a model system with two interacting electrons under the static ionic soft-Coulomb potential:
\begin{equation}
    v_{\text{ion}}(\mathbf{r}) = v_{\text{ion}}(x) = -\frac{Z_1}{\sqrt{(x-\frac{d}{2})^2+a^2}} - \frac{Z_2}{\sqrt{(x+\frac{d}{2})^2+a^2}},
\end{equation}
where $Z_1$ and $Z_2$ denote the charge of the atoms, $d$ denotes the bond length, and $a$ is a softening parameter for numerical stability.

The system is excited with a laser given by a time-dependent external potential:
\begin{equation}
    v_{\text{las}}(t) = A\sin{\omega t}.
\end{equation}

Here we assume the dipole approximation, where we can treat the laser as spatially constant because the size of our molecule is much smaller than the wavelength of the laser. Given the ground state prepared with $v_{\text{ion}}(x)$, we propagate the density under the influence of the laser $v_{\text{las}}(t)$.

For the electron-electron interaction we also consider a soft-Coulomb potential
\begin{equation}
    v_{\text{ee}}(\mathbf{r}, \mathbf{r'}) = v_{\text{ee}}(x,x') = \frac{1}{\sqrt{(x-x')^2+a^2}}\ .
\end{equation}
Within the ALDA, used here for the TDDFT calculations, we adopt the one-dimensional forms of the exchange~\cite{1D_LDA_Exchange} and correlation~\cite{1D_LDA_Correlation} functionals as implemented in the Octopus real-time TDDFT code~\cite{tancogne-dejean_octopus_2020}. For both the external potential and the electron–electron interaction, the softening parameter of the soft-Coulomb interaction is chosen as $a = 1$.

The spatial domain is defined by [-L, L] for time [0, T] with discretization $\Delta x$ and $\Delta t$, and we use fixed boundary conditions
\begin{equation}
    \phi_i(-L, t) = \phi_i(L, t) = 0, \quad \forall t \in [0, T].
\end{equation}

% The ground state density, ground state potential terms and shape of the laser pulse are shown in in Figure~\ref{fig:example_system}. 
We use the LDA functional due to its simplicity and accuracy for such simple one-dimensional systems. The density evolution is calculated by solving the time-dependent Kohn-Sham equations with the external potential $v_{\text{ext}}(x, t)$ using the Crank-Nicholson scheme. The numerical simulation is performed using the Octopus real time TDDFT code \cite{tancogne-dejean_octopus_2020}.

\subsection{Physics-Informed Machine Learning}
\begin{figure*}[!ht]
    \centering
    \includegraphics[width=0.9\textwidth]{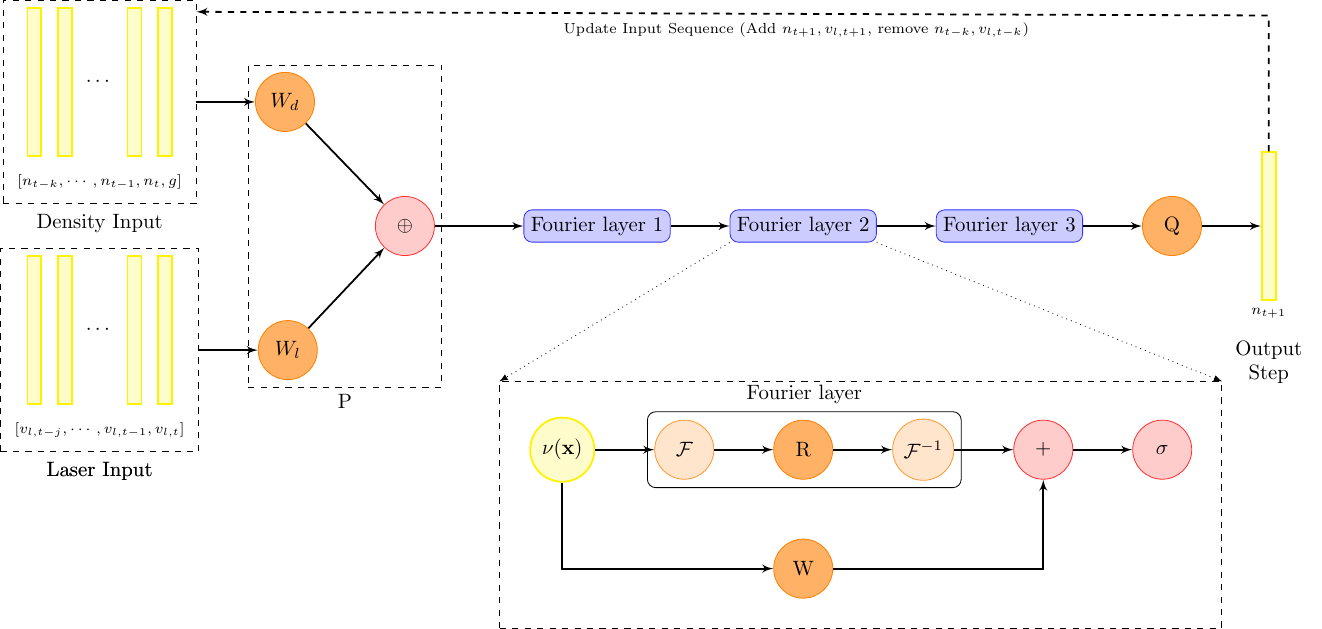}
    \caption{Autoregressive FNO architecture for predicting the density at time $t+1$ based on $k$ previous time steps.}
    \label{fig:FNO_arch}
\end{figure*}

Neural operators (NOs) \cite{luDeepONetLearningNonlinear2019, anandkumarNeuralOperatorGraph2020} extend neural networks by mapping functions to functions instead of finite-dimensional vectors. While a neural network maps $ \mathbb{R}^n $ to $ \mathbb{R}^m $, a neural operator maps function spaces, $ \mathcal{G}: \mathcal{A} \rightarrow \mathcal{U} $. Our goal is to approximate the non-linear map $\mathcal{G}^{\dagger}: \mathcal{A} \rightarrow \mathcal{U}$ with a neural operator $\mathcal{G}_\theta$, parameterized by $\theta \in \mathbb{R}^p$.
Training involves observations $ \{ (a_i, u_i) \}_{i=1}^N $ where $ u_i = \mathcal{G}^{\dagger}(a_i) $. The objective is to find parameters $\theta^*$ minimizing the loss:
\begin{equation}
\theta^* = \min_{\theta \in \mathbb{R}^p} \frac{1}{N} \sum_{i=1}^N \left\| u_i - \mathcal{G}_\theta(a_i) \right\|_{\mathcal{U}}^2.
\end{equation}
A neural operator $\mathcal{G}_\theta(a)$ is defined by the composition of layers (see also Figure~\ref{fig:FNO_arch}):
\begin{equation}
\mathcal{G}_\theta(a) = Q \circ f^{(J)} \circ f^{(J-1)} \circ \dots \circ f_1 \circ P (a),
\end{equation}
where $P$ maps the input function $a$ to a higher dimensional representation $f_0(x)$, $Q$ maps the final layer's output $f^{(J)}(x)$ back to the target function space $\mathcal{U}$. This mapping is done pointwise, projecting the vector at each spatial point into a vector of higher dimension to provide a richer latent representation for the model. Each intermediate layer $f^{(j+1)}$ updates the representation $f^{(j)}$ via:
\begin{equation}
f_{j+1}(x) = \sigma \left( W^{(j)} f^{(j)}(x) + (\mathcal{K}^{(j)}(a; \psi^{(j)}) f^{(j)})(x) \right), \quad \forall x \in D.
\end{equation}
Here, $W^{(j)}$ is a learnable weights matrix which corresponds to a linear transformation, $\sigma$ is a non-linear activation function applied point-wise, and $\mathcal{K}^{(j)}$ is a non-local kernel integral operator parameterized by $\psi^{(j)}$:
\begin{equation}
(\mathcal{K}^{(j)}(a; \psi^{(j)}) f^{(j)})(x) = \int_{D} \kappa^{(j)}(x, y, a(x), a(y); \psi^{(j)}) f^{(j)}(y) \, dy.
\end{equation}
The kernel function $\kappa^{(j)}$ integrates information across the domain $D$. The parameters $\theta$ consist of the parameters of $P$, $Q$, and all $W^{(j)}$ and $\psi^{(j)}$.

The Fourier Neural Operator (FNO) \cite{liFourierNeuralOperator2020} proposes an efficient implementation of the kernel integral operator $\mathcal{K}^{(j)}$. The FNO restricts the kernel integration to be a convolution operator, assuming the kernel depends only on the displacement, i.e., $\kappa^{(j)}(x, y; \psi^{(j)}) = \kappa^{(j)}(x-y; \psi^{(j)})$. This allows the use of the Convolution Theorem, which states that a convolution in the spatial domain is equivalent to point-wise multiplication in the Fourier domain:
\begin{equation}
\label{eq:conv_theorem}
(\mathcal{K}^{(j)} f^{(j)})(x) = (\kappa^{(j)} * f^{(j)})(x) = \mathcal{F}^{-1} \left( \mathcal{F}(\kappa^{(j)}) \cdot \mathcal{F}(f^{(j)}) \right)(x),
\end{equation}
where $\mathcal{F}$ denotes the Fourier transform and $\mathcal{F}^{-1}$ its inverse. The operator $\cdot$ represents point-wise multiplication in the Fourier domain.

Instead of learning the kernel $\kappa^{(j)}$ directly in the spatial domain, the FNO parameterizes the operator directly in the Fourier domain. The Fourier coefficients $\mathcal{F}(\kappa^{(j)})$ are represented by a complex-valued tensor $R_{\psi^{(j)}}$ which constitutes the learnable parameters $\psi^{(j)}$ for the kernel operator at layer $j$. The FNO truncates the Fourier series, considering only a finite number of lower-frequency modes, $k \in \{ k \mid \|k\| \le k^{(j)}_{\text{max}} \}$, where $k^{(j)}_{\text{max}}$ is the maximum frequency mode for layer $j$. This significantly reduces the number of parameters. The operation in \Eqref{eq:conv_theorem} is thus implemented as:
\begin{enumerate}
    \item Compute the Fourier transform of the input $f^{(j)}(x)$, yielding $\mathcal{F}(f^{(j)})(k)$.
    \item For each Fourier mode $k$ up to $k^{(j)}_{\text{max}}$, perform a linear transformation (element-wise multiplication) with the corresponding parameters in $R_{\psi^{(j)}}$: $(\widehat{f^{(j)}})_k = (R_{\psi^{(j)}})_k \cdot (\mathcal{F}(f^{(j)}))_k$.
    \item Compute the inverse Fourier transform of the result: $\mathcal{F}^{-1}(\widehat{f^{(j)}})(x)$.
\end{enumerate}
The update rule for the FNO layer becomes:
\begin{equation}
f_{j+1}(x) = \sigma \left( W^{(j)} f^{(j)}(x) + \mathcal{F}^{-1} \left( R_{\psi^{(j)}} \cdot (\mathcal{F}(f^{(j)})) \right)(x) \right).
\end{equation}
This formulation has several advantages. The Fast Fourier Transform (FFT) algorithm allows computing $\mathcal{F}$ and $\mathcal{F}^{-1}$ efficiently, typically in $\mathcal{O}(n \log n)$ time, where $n$ is the number of discretization points. The parameterization in the Fourier domain makes the FNO discretization-invariant, meaning the same learned operator can be evaluated on different mesh resolutions without retraining. The number of modes $k_{\text{max}}$ and the width of the channel dimensions (determined by $Q,P$ and $W^{(j)}$) control the expressivity and complexity of the model. By stacking these layers, FNOs can effectively capture complex, non-linear dependencies between functions while maintaining computational efficiency.

\subsection{Time-Propagator Design} \label{FNO_time}

To model the evolution of the electron density $n(x, t)$ under the influence of a time-dependent external potential $v_l(t)$, we employ an autoregressive approach based on FNOs whose architecture is well-suited for learning resolution-invariant mappings between infinite-dimensional function spaces, making it appropriate for processing spatio-temporal data like the electron density \cite{liFourierNeuralOperator2020}. The autoregressive framework allows the model to predict future states based on a sequence of past observations and the corresponding external potential values.

Let $\mathbf{n}_t \in \mathbb{R}^{n_x}$ denote the discretized electron density on a spatial grid of size $n_x$ at time $t$, and let $v_{l,t} \in \mathbb{R}$ be the external potential value. The input representation module $P$ maps these inputs to a higher-dimensional feature space suitable for the FNO layers.

We define the density input sequence as:
\begin{equation}
\mathbf{N}_t = [\mathbf{n}_{t-T_{\text{in}}+1}, \ldots, \mathbf{n}_t] \in \mathbb{R}^{n_x \times T_{\text{in}}},
\end{equation}
and the potential sequence as:
\begin{equation}
\mathbf{V}_{l,t} = [v_{l,t-T_{\text{in,laser}}+1}, \ldots, v_{l,t}] \in \mathbb{R}^{n_x \times T_{\text{in,laser}}},
\end{equation}
where $T_{\text{in}}$ and $T_{\text{in,laser}}$ denote the number of past time steps for density and potential inputs, respectively. Additionally, we optionally include the spatial grid coordinates $\mathbf{g} \in \mathbb{R}^{n_x \times 1}$.

The initial input sequences, $\mathbf{N}_0$ and $\mathbf{V}_{l,0}$, consist of the first $T_{\text{in}}$ time steps, starting with the ground state density $\mathbf{n}_{0}$ and the initial potential values.

The FNO model, denoted by $\mathcal{G}_\theta$ with trainable parameters $\theta$, maps the input sequences to the predicted density at the next time step, $\hat{\mathbf{n}}_{t+1}$:
\begin{equation}
\hat{\mathbf{n}}_{t+1} = \mathcal{G}_\theta(\mathbf{N}_t, \mathbf{V}_{l,t}).
\end{equation}
Here, the model implicitly learns time propagation through the recent history of the density, the applied potential, and the subsequent density state.

Following the prediction of $\hat{\mathbf{n}}_{t+1}$, the input sequences are updated for the next prediction step ($t+1 \rightarrow t+2$). The new density sequence $\mathbf{N}_{t+1}$ is formed by appending the predicted density $\hat{\mathbf{n}}_{t+1}$ and removing the oldest density grid $\mathbf{n}_{t-T_{\text{in}}+1}$:
\begin{equation}
\mathbf{N}_{t+1} = \left[ \mathbf{n}_{t-T_{\text{in}}+2}, \mathbf{n}_{t-T_{\text{in}}+3}, \ldots, \mathbf{n}_{t}, \hat{\mathbf{n}}_{t+1} \right].
\end{equation}
Similarly, the potential sequence is updated by incorporating the next known potential value $v_{l, t+1}$ and removing the oldest value:
\begin{equation}
\mathbf{V}_{l,t+1} = \left[ v_{l, t-T_{\text{in}}+2}, v_{l, t-T_{\text{in}}+3}, \ldots, v_{l, t}, v_{l, t+1} \right].
\end{equation}
This autoregressive process, illustrated in Figure~\ref{fig:FNO_arch}, is repeated iteratively to generate the density evolution over the desired time period.

\subsubsection{Input Representation Design}

The autoregressive FNO framework requires an effective representation of both the electron density history and the time-dependent external potential. The representation consists of density grids $\mathbf{N}$, external potential grids $\mathbf{V}$ and a normalized spatial grid $\mathbf{g}$ with coordinate points in the range [-1,1], providing explicit positional information. We implement the following input representation strategies to capture the complex spatio-temporal dependencies between these quantities:

\textbf{1. Density-Only Representation:} This baseline approach uses only the density history:
\begin{equation}
P_{\text{density}}(\mathbf{N}_t, \mathbf{g}) = W_d [\mathbf{N}_t, \mathbf{g}],
\end{equation}
where $W_d \in \mathbb{R}^{w \times (T_{\text{in}}+1)}$ maps to the model width $w$, and $[\cdot, \cdot]$ denotes concatenation along the feature dimension.

\textbf{2. Direct Concatenation:} A straightforward approach that concatenates all inputs:
\begin{equation}
P_{\text{concat}}(\mathbf{N}_t, \mathbf{V}_{l,t}, \mathbf{g}) = W_c [\mathbf{N}_t, \mathbf{V}_{l,t}, \mathbf{g}],
\end{equation}
where $W_c \in \mathbb{R}^{w \times (T_{\text{in}}+T_{\text{in,laser}}+1)}$.

\textbf{3. Separate Processing with Concatenation:} This approach processes density and potential through independent transformations before concatenation:
\begin{equation}
P_{\text{sep-concat}}(\mathbf{N}_t, \mathbf{V}_{l,t}, \mathbf{g}) = [W_d [\mathbf{N}_t, \mathbf{g}], W_l [\mathbf{V}_{l,t}]],
\end{equation}
where $W_d \in \mathbb{R}^{w \times (T_{\text{in}}+1)}$ and $W_l \in \mathbb{R}^{w \times T_{\text{in,laser}}}$, yielding an output dimension of $2w$. The transformation is implemented as separate fully connected layers with weight matrices, $W_d$ for the density input and $W_l$ for the laser input. Both matrices are trainable parameters optimized during training.

\textbf{4. Separate Processing with Addition:} The default representation processes inputs independently and combines them additively:
\begin{equation}
P_{\text{sep-add}}(\mathbf{N}_t, \mathbf{V}_{l,t}, \mathbf{g}) = W_d [\mathbf{N}_t, \mathbf{g}] + W_l [\mathbf{V}_{l,t}],
\end{equation}
maintaining the output dimension $w$.

\textbf{5. Space-Time Convolution:} This approach treats the input as a 2D spatio-temporal field. The density and potential sequences are reshaped and stacked:
\begin{equation}
    \mathbf{X} =
    \left[
    \begin{array}{c}
    \mathbf{N}_t^T \\ 
    \mathbf{V}_{l,t}^T
    \end{array}
    \right]
    \in \mathbb{R}^{2 \times T_{\text{in}} \times n_x}.
    \end{equation}
where we pad $\mathbf{V}_{l,t}$ to match $T_{\text{in}}$ if necessary. A 2D convolutional operator with kernel $\mathcal{K}_{\text{st}} \in \mathbb{R}^{w \times 2 \times k_t \times k_x}$ is applied:
\begin{equation}
P_{\text{st-conv}}(\mathbf{N}_t, \mathbf{V}_{l,t}, \mathbf{g}) = \text{AvgPool}_t(\mathcal{K}_{\text{st}} * \mathbf{X}) + W_g \mathbf{g},
\end{equation}
where $*$ denotes 2D convolution, $\text{AvgPool}_t$ performs temporal averaging, and $W_g \in \mathbb{R}^{w \times 1}$ incorporates spatial grid information.

The choice of input representation affects the model's ability to capture different types of dependencies. The separate processing approaches (3 and 4) allow the model to learn distinct features for density and potential before combination, while the space-time convolution explicitly models spatio-temporal correlations through convolutional filters. The output of the input representation module $P$ serves as the initial feature representation $f^{(0)}(x)$ for the subsequent FNO layers.

\subsubsection{Observable Calculations}

To evaluate the physical validity and accuracy of the predicted electron densities, we compute several key observables from both the predicted and reference density fields.

\textbf{Integrated Density (Particle Number):} The total number of electrons is computed by integrating the density over the spatial domain:
\begin{equation}
N(t) = \int_{-\infty}^{\infty} n(x,t) \, dx \approx \sum_{i=1}^{n_x} n(x_i, t) \, \Delta x,
\end{equation}
where $n_x$ is the number of spatial grid points and $\Delta x$ is the grid spacing. For our systems, this integral should remain constant at $N = 2$ throughout the evolution, serving as a fundamental conservation check.

\textbf{Dipole Moment:} The dipole moment characterizes the charge distribution asymmetry and is particularly sensitive to the system's response to the external field:
\begin{equation}
\mu(t) = \int_{-\infty}^{\infty} x \, n(x,t) \, dx \approx \sum_{i=1}^{n_x} x_i \, n(x_i, t) \, \Delta x.
\end{equation}
The time-dependent dipole moment directly relates to the optical response and provides insight into the collective electron dynamics.

\textbf{Thomas-Fermi Total Energy:} We cannot evaluate the Kohn-Sham total energy as the FNO model predicts the total electron density while we need the Kohn-Sham orbitals to evaluate the Kohn-Sham kinetic energy. Instead we use the Thomas-Fermi approximation \cite{parr_density-functional_1995, martin_electronic_2004} as a proxy for the total energy functional which is defined as the sum of kinetic, Hartree, and external potential contributions. The kinetic energy is approximated as:
\begin{equation}
E_K[n] = \frac{\pi^2}{6} \int_{-\infty}^{\infty} n(x,t)^3 \, dx \approx \frac{\pi^2}{6} \sum_{i=1}^{n_x} n(x_i, t)^3 \, \Delta x.
\end{equation}

The Hartree energy accounts for electron-electron repulsion:
\begin{equation}
E_H[n] = \frac{1}{2} \int_{-\infty}^{\infty} \int_{-\infty}^{\infty} \frac{n(x,t) n(x',t)}{|x-x'|} \, dx \, dx',
\end{equation}
which we evaluate numerically using a soft-Coulomb regularization:
\begin{equation}
E_H[n] \approx \frac{1}{2} \sum_{i,j=1}^{n_x} \frac{n(x_i,t) n(x_j,t)}{\sqrt{(x_i-x_j)^2 + a^2}} \, (\Delta x)^2,
\end{equation}
where $a$ is a small regularization parameter.

The external potential energy includes contributions from both the ionic potential $v_{\text{ion}}(x)$ and the time-dependent laser field $v_l(t)$:
\begin{equation}
E_{\text{ext}}[n] = \int_{-\infty}^{\infty} n(x,t) [v_{\text{ion}}(x) + v_l(t)] \, dx.
\end{equation}

The total energy in the Thomas-Fermi approximation is then:
\begin{equation}
E_{\text{tot}}[n] = E_K[n] + E_H[n] + E_{\text{ext}}[n].
\end{equation}

These observables are computed at each time step for both predicted and reference densities. The absolute errors in these quantities provide quantitative measures of the model's physical accuracy beyond point-wise density comparisons. The dipole moment error is particularly important as it directly relates to effects of the time-dependent laser field, while energy conservation provides a  test of the model's ability to capture the system's dynamics correctly.

\subsubsection{Physics-Informed Loss Functions}

To enhance the physical consistency of the predictions, we incorporate physics-informed loss terms alongside the standard data-driven loss. The combined loss function takes the form:
\begin{equation}
\mathcal{L}_{\text{total}} = \mathcal{L}_{\text{data}} + \lambda \mathcal{L}_{\text{int}}
% + \lambda_2 \mathcal{L}_{\text{dipole}} + \lambda_3 \mathcal{L}_{\text{energy}},
\end{equation}
where $\lambda$ is a hyperparameter that controls the relative importance of the integral constraint.

\textbf{Data-Driven Loss:} The primary loss term measures the point-wise discrepancy between predicted and reference densities:
\begin{equation}
\mathcal{L}_{\text{data}} = \frac{1}{|\mathcal{D}|} \sum_{d \in \mathcal{D}} \frac{1}{|\mathcal{T}|} \sum_{t \in \mathcal{T}} \left\| \mathbf{n}_{t}^{(d)} - \hat{\mathbf{n}}_{t}^{(d)} \right\|_2^2,
\end{equation}
where $\mathcal{D}$ denotes the training dataset and $\mathcal{T}$ the set of predicted time steps.

\textbf{Integral Conservation Loss:} This term enforces particle number conservation:
\begin{equation}
\mathcal{L}_{\text{int}} = \frac{1}{|\mathcal{D}|} \sum_{d \in \mathcal{D}} \frac{1}{|\mathcal{T}|} \sum_{t \in \mathcal{T}} \left( \sum_{i} \hat{n}_{t,i}^{(d)} \Delta x - N_{\text{ref}}^{(d)} \right)^2,
\end{equation}
where $N_{\text{ref}}^{(d)} = 2$ is the reference number of electrons for system $d$.

% \textbf{Dipole Moment Loss:} To ensure accurate optical response, we penalize deviations in the dipole moment:
% \begin{equation}
% \mathcal{L}_{\text{dipole}} = \frac{1}{|\mathcal{D}|} \sum_{d \in \mathcal{D}} \frac{1}{|\mathcal{T}|} \sum_{t \in \mathcal{T}} \left| \mu_{t}^{(d)} - \hat{\mu}_{t}^{(d)} \right|^2,
% \end{equation}
% where $\mu_{t}^{(d)}$ and $\hat{\mu}_{t}^{(d)}$ are the reference and predicted dipole moments, respectively.

% \textbf{Total Energy Loss:} This term promotes energy consistency throughout the evolution:
% \begin{equation}
% \mathcal{L}_{\text{energy}} = \frac{1}{|\mathcal{D}|} \sum_{d \in \mathcal{D}} \frac{1}{|\mathcal{T}|} \sum_{t \in \mathcal{T}} \left| E_{\text{tot}}[n_{t}^{(d)}] - E_{\text{tot}}[\hat{n}_{t}^{(d)}] \right|^2,
% \end{equation}
% where the total energy is computed using the Thomas-Fermi functional described above.

% We systematically study the effects of different loss term combinations by varying the weights $\lambda_1$, $\lambda_2$, and $\lambda_3$. This allows us to investigate the trade-offs between point-wise accuracy and physical consistency, and to determine the optimal balance for different prediction scenarios. 
The inclusion of physics-informed losses not only improves the physical plausibility of the predictions but also enhances the model's generalization to systems outside the training distribution.
Furthermore, we include this loss term in all models below, as inclusion of this term stabilizes training and makes models reach physical consistency faster. 

\subsubsection{Error Metrics}

To evaluate model performance, we employ multiple error metrics that capture different aspects of prediction accuracy for both point-wise density fields and integrated observables. Here, $y_i$ and $\hat{y}_i$ denote reference and predicted values respectively.

\textbf{Mean Squared Error (MSE):} The MSE quantifies the average squared deviation between predicted and reference values:
\begin{equation}
\text{MSE} = \frac{1}{n} \sum_{i=1}^{n} (y_i - \hat{y}_i)^2,
\end{equation}
For density fields, this is computed point-wise across all spatial locations and time steps.

\textbf{Mean Absolute Error (MAE):} The MAE provides a more interpretable measure of average prediction error:
\begin{equation}
\text{MAE} = \frac{1}{n} \sum_{i=1}^{n} |y_i - \hat{y}_i|.
\end{equation}
Unlike MSE, MAE is less sensitive to outliers and provides errors in the same units as the predicted quantity.

\textbf{Mean Absolute Percentage Error (MAPE):} For observables with varying magnitudes, MAPE normalizes the error relative to the reference value:
\begin{equation}
\text{MAPE} = \frac{100}{n} \sum_{i=1}^{n} \left| \frac{y_i - \hat{y}_i}{y_i} \right|,
\end{equation}
where we apply a small tolerance $\epsilon = 10^{-6}$ to avoid division by zero: $\max(|y_i|, \epsilon)$.

\textbf{Symmetric Mean Absolute Percentage Error (SMAPE):} To address the asymmetry in MAPE, we also compute:
\begin{equation}
\text{SMAPE} = \frac{100}{n} \sum_{i=1}^{n} \frac{|y_i - \hat{y}_i|}{(|y_i| + |\hat{y}_i|)/2 + \epsilon}.
\end{equation}
SMAPE provides a bounded percentage error that treats over- and under-predictions symmetrically.

For density fields, we primarily report MSE and MAE computed point-wise across the spatial grid and averaged over time. For observables, particularly the dipole moment where values can approach zero, we employ all four metrics to provide a comprehensive assessment. The percentage-based metrics (MAPE and SMAPE) are especially valuable for the dipole moment, as its magnitude varies significantly during the laser pulse interaction, making absolute errors less informative. Visualizations, like those shown in Figure~\ref{fig:example_snapshot}, comparing the predicted and reference density evolution provide qualitative assessment.

\section{Results}
\label{sec:results}

\subsection{Dataset and Computational Setup}

We evaluate our autoregressive FNO model on a comprehensive dataset of TDDFT simulations. The dataset comprises 2048 independent simulations of diatomic systems with varying nuclear charges $(Z_1, Z_2)$, internuclear distances $d$, and laser field parameters including intensity and wavelength. Each simulation is performed on a spatial grid of 361 points with spacing $\Delta x = 0.05$ a.u., covering the domain $[-9.0, 9.0]$ a.u. The temporal evolution spans $[0.0, 5.0]$ fs with 51 time steps at $\Delta t = 0.1$ fs resolution. 

The reference data is generated using the Octopus real-space TDDFT code with a fine temporal resolution of $\Delta t = 0.01$ fs to ensure numerical accuracy. For consistent comparison with our model predictions, we also compute reference solutions at the coarser $\Delta t = 0.1$ fs resolution using the same TDDFT solver. This allows us to demonstrate the accuracy of the model with larger time steps. We filter the dataset to remove systems with low time-dependent variation and systems with numerical artifacts such as boundary reflections. The final dataset consists of 800 training systems, 150 validation systems and 200 test systems representing various input conditions and laser parameters. For long-horizon stability, we additionally roll out to $\SI{10.0}{fs}$ without retraining and evaluate on a held-out subset. We enumerate the parameters used for Octopus data generation in the \hyperref[sec:SI]{SI}.

\subsection{Baseline Performance and Ablation Studies}

\begin{figure}[h!]
    \centering
    \includegraphics[width=0.8\textwidth]{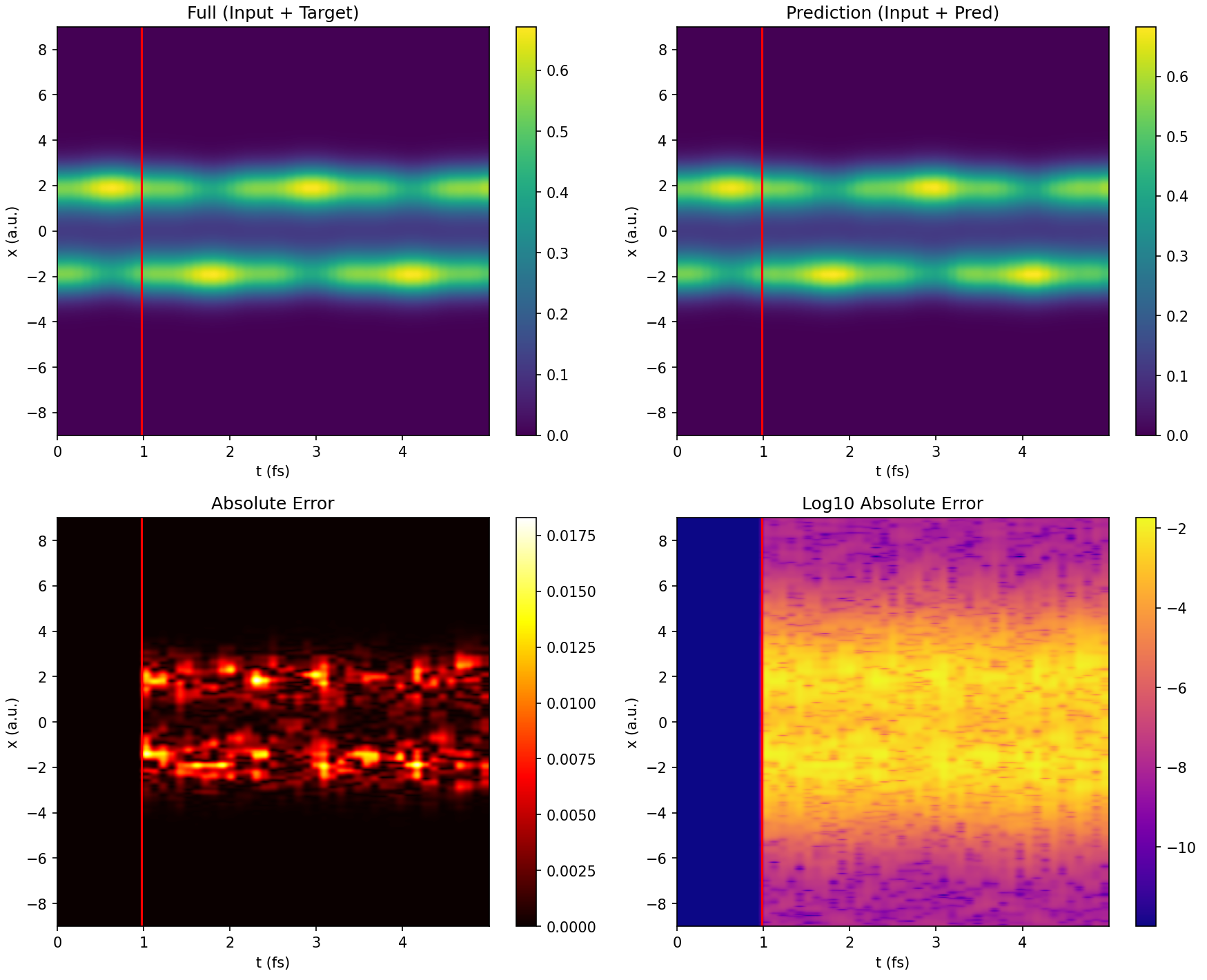}
    \caption{Results for a representative system.}
    \label{fig:example_snapshot}
\end{figure}

We first establish baseline performance by comparing different input representation strategies and conducting ablation studies on key architectural components. Figure~\ref{fig:example_snapshot} shows a representative rollout. The red vertical line marks the boundary between the $T_{\text{in}}$ input slices and the predicted slices. Using the fully connected input encoding for densities and lasers, the rollout remains stable and tracks fine-resolution real time TDDFT closely.
Table~\ref{tab:input_representation} compares input encodings. The results demonstrate that incorporating laser information significantly improves prediction accuracy compared to the density-only baseline. Relative to the density-only baseline, the best variant reduces density MAE from $1.97\times 10^{-3}$ to $0.872\times 10^{-3}$ and dipole MSE from $11.4\times 10^{-3}$ to $7.37\times 10^{-3}$, while keeping inference time at the millisecond level per step. The two-stream encodings that process density and laser separately and combine them by addition or concatenation outperform density-only and pure spacetime-convolution variants. All learned variants keep the integrated density very close to the target value of 2.
\begin{table}[h!]
    \centering
    \caption{Comparison of error metrics across input representations.}
    \begin{tabular}{lccccc}
    \toprule
    Metric & Density Only & Concat &  FC Add &  FC Concat & Spacetime Conv \\
    \midrule
    AE ($10^{-3}$)         & 1.97  & 0.899 & \textbf{0.872} & 0.879 & 1.48 \\
    MSE ($10^{-5}$)        & 6.15  & \textbf{3.13} & 3.20 & 3.27 & 5.08 \\
    MAPE (\%)              & 3.71  & 1.90  & \textbf{1.84} & 1.85 & 3.80 \\
    Dipole MSE ($10^{-3}$) & 11.4  & \textbf{6.75} & 7.37 & 7.27 & 8.92 \\
    Dipole MAPE (\%)       & 27.6  & 18.6  & \textbf{9.35} & 10.7 & 27.0 \\
    Integral ($\approx 2$) & 1.9990 & \textbf{2.0000} & 1.9995 & \textbf{2.0000} & 2.0005 \\
    Inference Time (ms)    & 1.33 & \textbf{1.32} & 1.35 & 1.38 & 1.51 \\
    \bottomrule
    \end{tabular}
    \label{tab:input_representation}
\end{table}

\begin{table}[h!]
        \centering
        \caption{Comparison of error metrics between ML time propagator and numerical simulation on the same grid.}
        \begin{tabular}{lcc}
        \toprule
        Metric & FNO & Octopus Coarse \\
        \midrule
        AE ($10^{-3}$)             & \textbf{0.872} & 6.10 \\
        MSE ($10^{-5}$)            & \textbf{3.20}  & 31.2 \\
        MAPE (\%)                  & \textbf{1.84}  & 853 \\
        SMAPE (\%)                 & \textbf{1.73}  & 62.3 \\
        Dipole MSE ($10^{-3}$)     & \textbf{7.37}  & 79.0 \\
        Dipole MAPE (\%)           & \textbf{9.35}  & 74.0 \\
        Dipole SMAPE (\%)          & \textbf{6.08}  & 51.1 \\
        \midrule
        Integral ($\approx 2$)     & 1.9995 $\pm$ 0.0014 & \textbf{2.0000 $\pm$ 2.0e-7} \\
        Inference Time (ms)        & \textbf{3.07 $\pm$ 0.05} & 23.94 $\pm$ 2.19 \\
        \bottomrule
        \end{tabular}
        \label{tab:coarse_comparison}
\end{table}

\begin{table}[h!]
      \centering
      \caption{Comparison of error metrics across Spatial Superresolution, Time Extension, and Time Reversed.}
      \begin{tabular}{lccc}
      \toprule
      Metric & Spatial Super-resolution & Time Extension & Time Reversed \\
      \midrule
      AE ($10^{-3}$)             & 0.942 & 0.896 & 1.053 \\
      MSE ($10^{-5}$)            & 3.30  & 2.53 & 3.68 \\
      MAPE (\%)                  & 2.06  & 2.65 & 3.30 \\
      SMAPE (\%)                 & 1.95  & 2.18 & 2.92 \\
      Dipole MSE ($10^{-3}$)     & 7.64  & 5.45 & 9.05 \\
      Dipole MAPE (\%)           & 13.10 & 10.03 & 4803.43 \\
      Dipole SMAPE (\%)          & 7.22  & 6.29 & 9.87 \\
      \midrule
      Integral ($\approx 2$)     & 1.9966 $\pm$ 0.0014 & 1.9995 $\pm$ 0.0010 & 1.9995 $\pm$ 0.0012 \\
      Calculation Time (ms)      & 3.37 $\pm$ 0.05 & 3.05 $\pm$ 0.04 & 3.02 $\pm$ 0.12 \\
      \bottomrule
      \end{tabular}
      \label{tab:generalizability}
\end{table}
We also compare against a numerical real time TDDFT baseline computed directly on the coarse grid. Table~\ref{tab:coarse_comparison} shows that the FNO substantially improves agreement with the high-resolution reference when both are evaluated at $\Delta t=\SI{0.1}{fs}$. Density MSE drops from $31.2\times 10^{-5}$ (Octopus coarse) to $3.20\times 10^{-5}$ (FNO), representing an almost $10\times$ decrease in MSE. Percentage errors for the dipole are also lower for FNO across MAPE and SMAPE. The large MAPE for Octopus coarse arises from divisions near dipole zero values. SMAPE provides a more robust comparison and still favors FNO. The time taken to calculate density at the next time step drops from 23.94 ms in the coarse Octopus calculation to 3.07 ms for the FNO density prediction, an $8\times$ decrease. 

\begin{figure}[h!]

    \centering
    \includegraphics[width=0.48\textwidth]{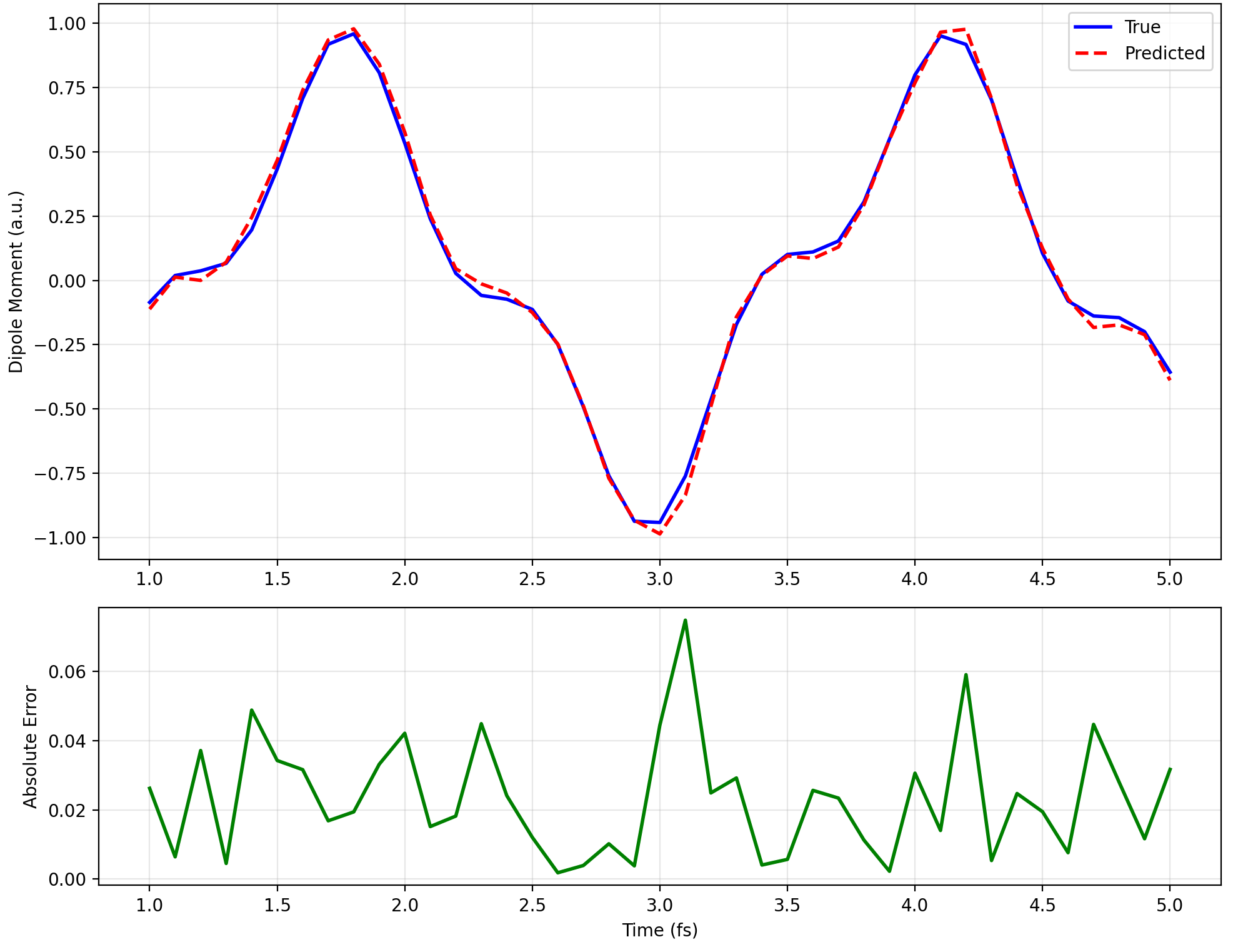}
    \includegraphics[width=0.48\textwidth]{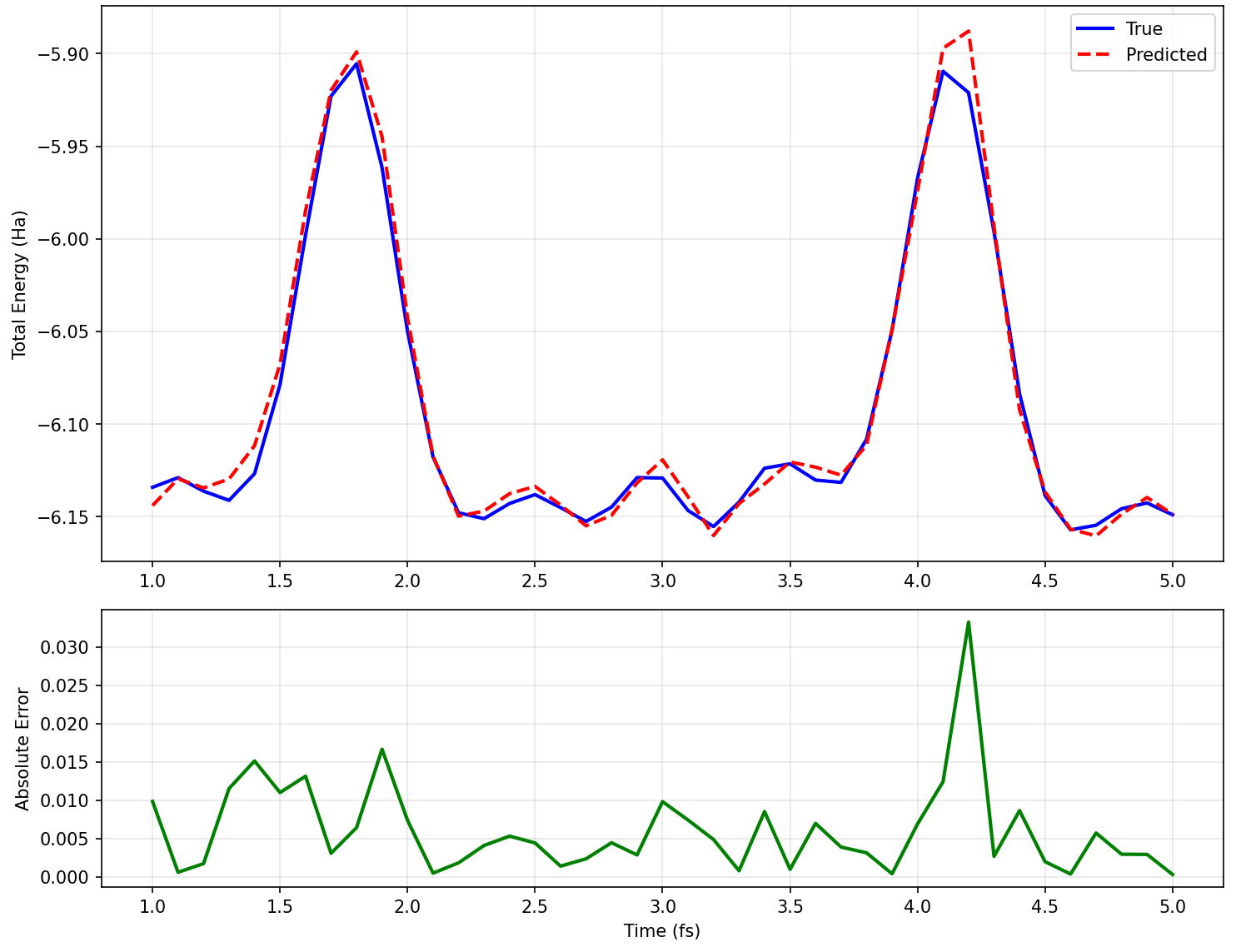}
    \caption{Observables calculated for the representative system. \\ 
Left: Dipole moment calculated from reference and predicted density. Right: Total TF energy calculated from reference and predicted density.}
    \label{fig:dipole_energy}
\end{figure}

\subsection{Generalization Capabilities}

To assess the model's ability to generalize beyond the training distribution, we evaluate its performance on two challenging scenarios: spatial super-resolution and extended temporal rollout.

\textbf{Spatial Super-resolution:} Evaluating the trained FNO at twice-finer spatial resolution ($\Delta x=\SI{0.025}{\au}$, 721 points) without retraining maintains errors consistent with the baseline model with a small increase in inference time (3.37 ms vs 3.07 ms) as shown in in Table~\ref{tab:generalizability}.

\textbf{Extended Temporal Rollout:}  Although training uses $[0,\SI{5.0}]$ fs, the model remains stable when rolled out to $\SI{10.0}{fs}$. In Table~\ref{tab:generalizability} the ``Time Extension'' column shows density MSE $\left(2.56\times 10^{-5}\right)$ and dipole MSE $\left(5.54\times 10^{-3}\right)$ at long horizon that are comparable to the base window. This indicates the autoregressive operator does not accumulate catastrophic error over the tested horizon.

\subsection{Observable Predictions}

Beyond point-wise density accuracy, we evaluate the model's ability to predict physically meaningful observables that are critical for understanding the system's dynamics.

\textbf{Dipole Moment:} Figure~\ref{fig:dipole_energy} compares predicted and reference dipole moments for representative systems. The model accurately captures the oscillatory behavior during laser excitation without error accumulation. 

\textbf{Thomas-Fermi Total Energy:} Within the Thomas-Fermi approximation, we compute the total energy functional comprising kinetic, Hartree, and external potential contributions. Figure~\ref{fig:dipole_energy} shows the temporal evolution of predicted versus reference total energies.

\subsection{Time-Propagator Properties}

We investigate whether the learned time-propagator exhibits fundamental physical properties expected from the underlying quantum dynamics.

\begin{figure}[h!]
    \centering
    \includegraphics[width=0.7\textwidth]{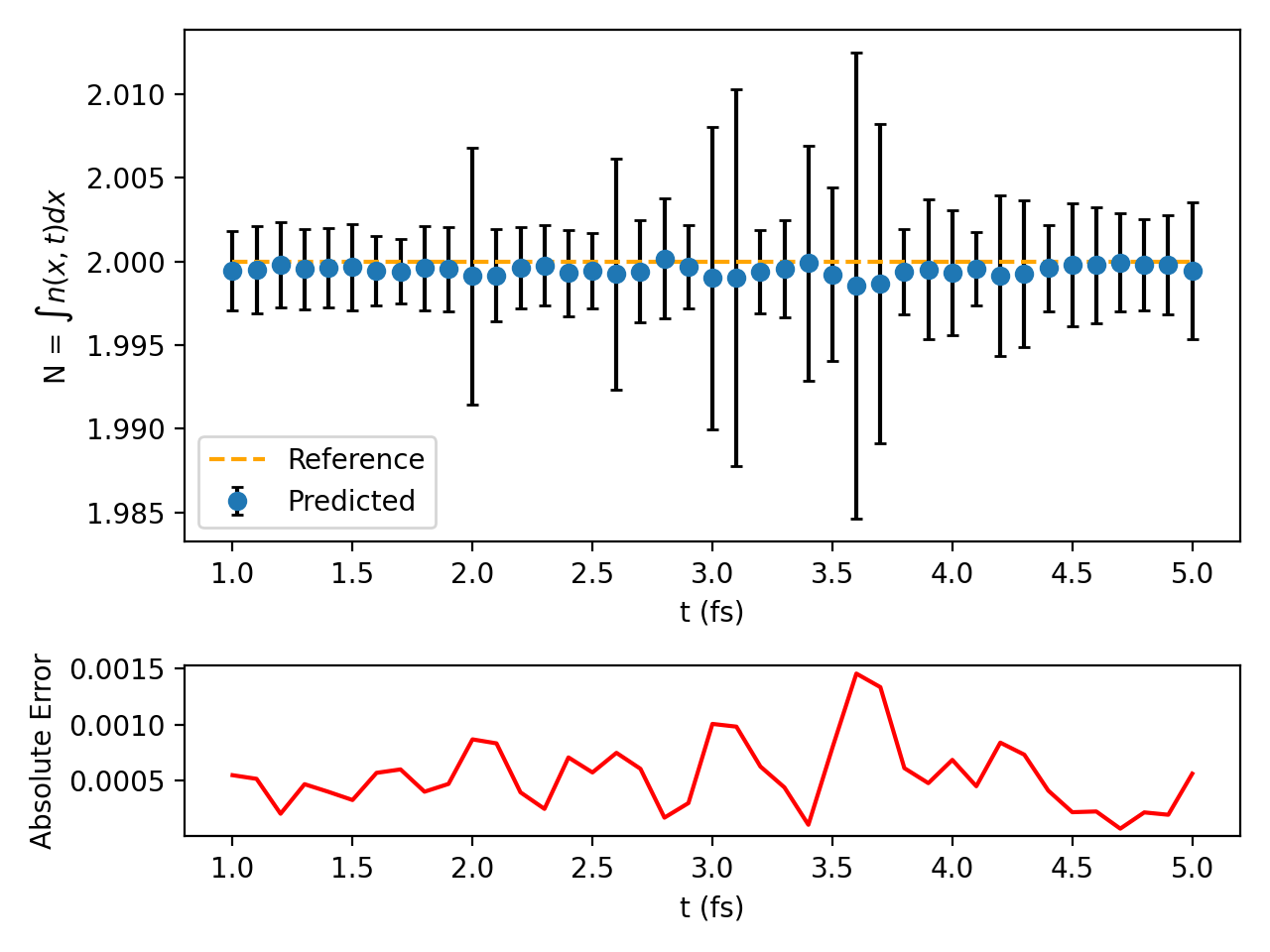}
    \caption{Integrated density over time for predicted rollouts. The conservation loss keeps particle number near 2 throughout.}
    \label{fig:integral_error}
\end{figure}

\textbf{Density Conservation:} fundamental requirement for any valid time evolution is the conservation of total particle number. Figure~\ref{fig:integral_error} aggregates the evolution of $\int n(x,t)\,dx$ across systems. With the integral loss, deviations remain within about $10^{-3}$ of the target value 2 over rollouts, preventing unphysical density gain or loss.

\begin{figure}[h!]
    \centering
    \includegraphics[width=0.8\textwidth]{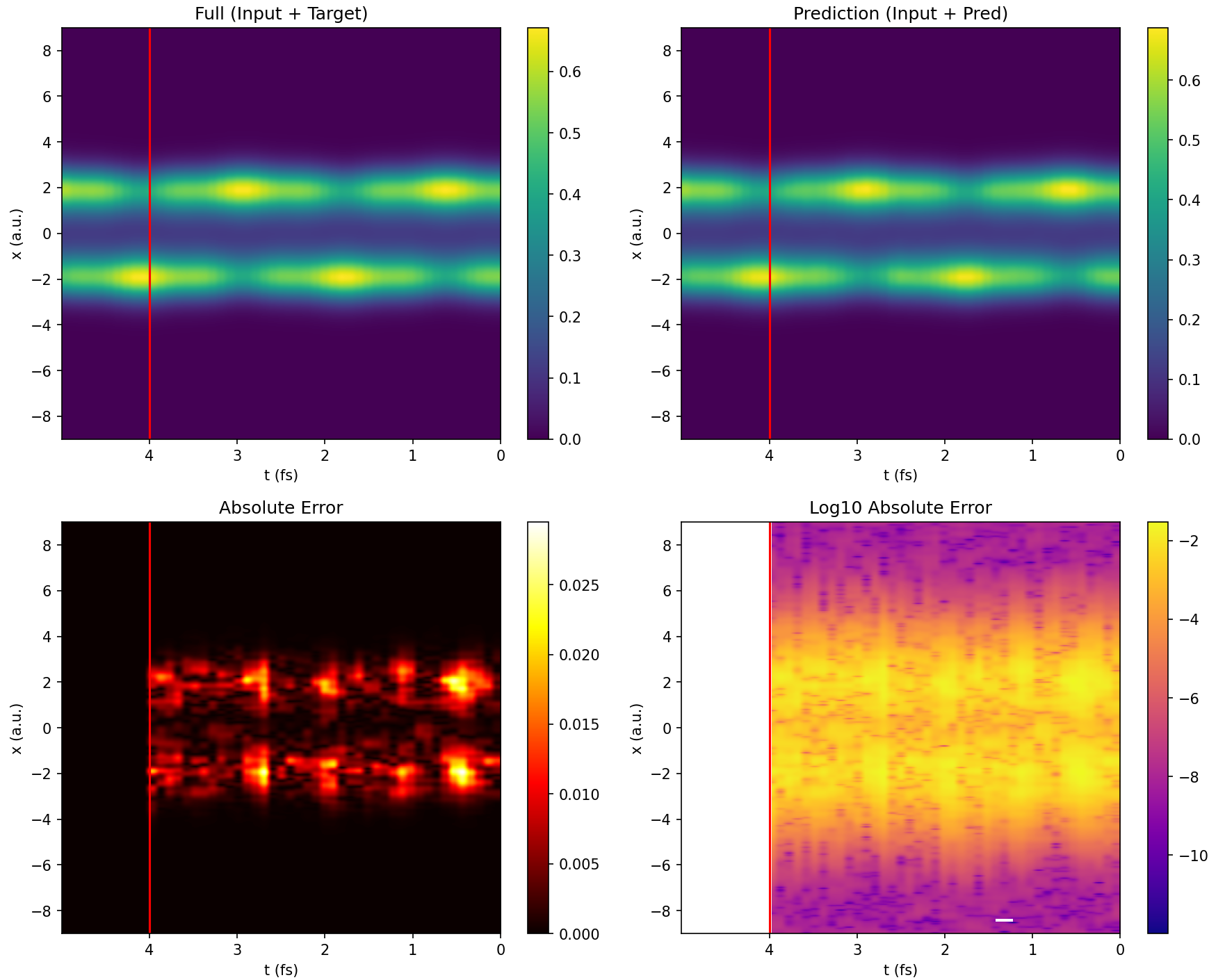}
    \caption{Results for the time reversed representative system.}
    \label{fig:tr_snapshot}
\end{figure}

\begin{figure}[h!]

    \centering
    \includegraphics[width=0.48\textwidth]{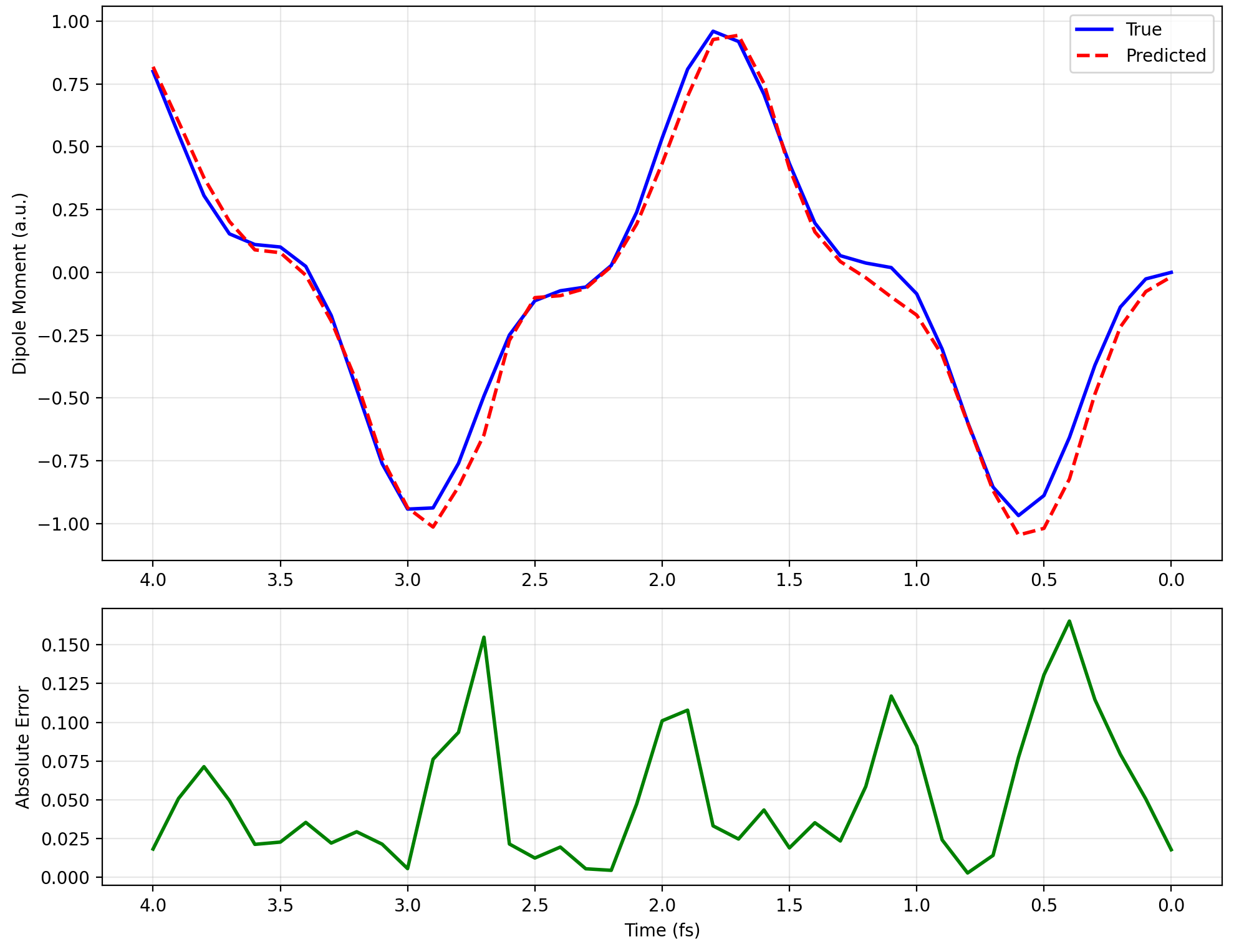}
    \includegraphics[width=0.48\textwidth]{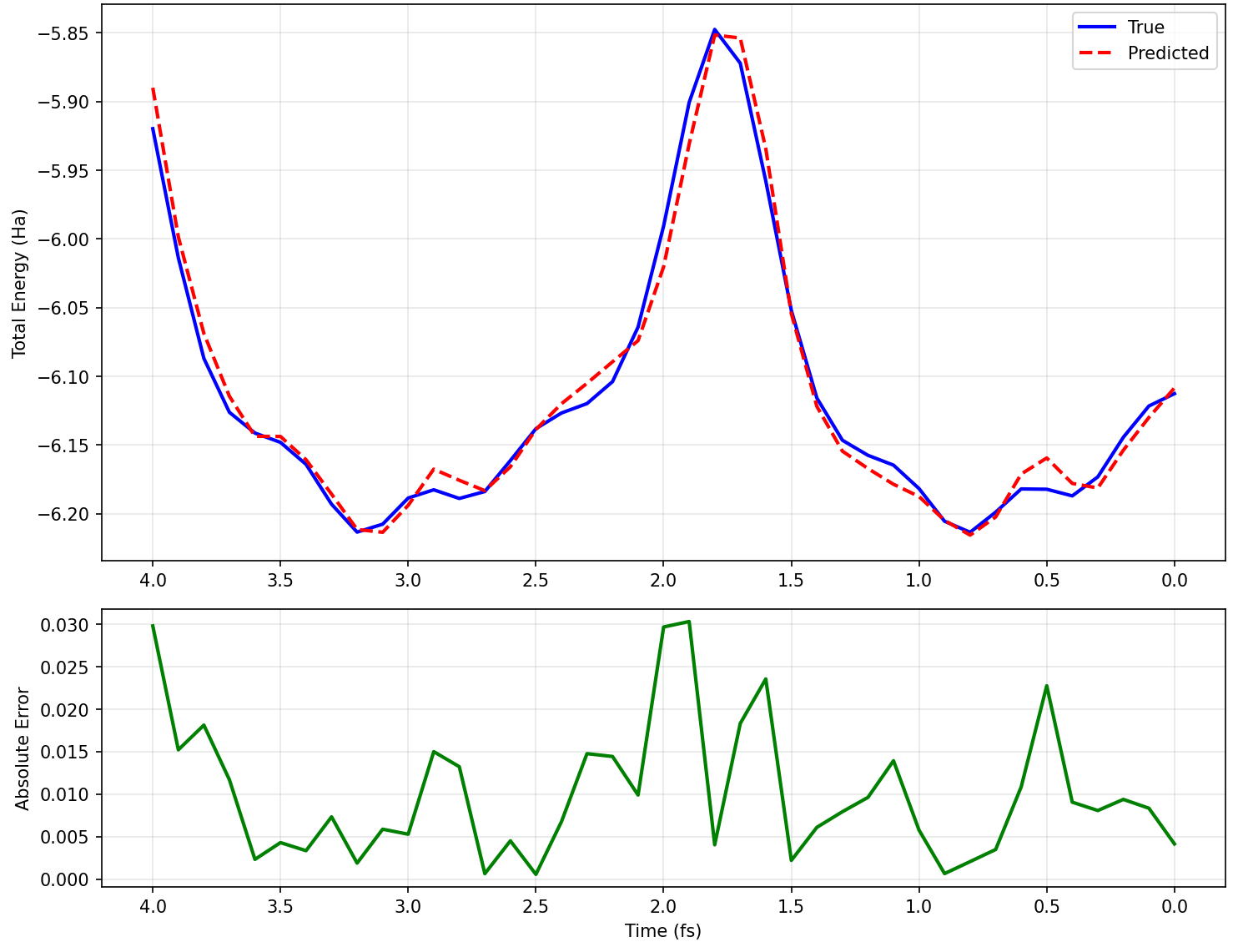}
    \caption{Observables calculated for the time reversed representative system. \\ 
Left: Dipole moment calculated from reference and predicted density. Right: Total TF energy calculated from reference and predicted density.}
    \label{fig:tr_dipole_energy}
\end{figure}

\textbf{Time Reversal Symmetry:} To test consistency with reversible dynamics, we start from states at $t=\SI{5.0}{fs}$ and propagate backward to $t=0$ using the time-reversed driving field. Figures~\ref{fig:tr_snapshot} and~\ref{fig:tr_dipole_energy} show the time reversed representative case, and metrics for all test systems are shown in Table~\ref{tab:generalizability}. The recovered densities remain close to the forward trajectory states at matching times, indicating that the trained model approximates a time-reversible operator in practice over the tested window. The error metrics are consistent with the baseline case, indicating that the time propagator exhibits approximate time reversal symmetry. The high MAPE can be accounted by the fact that at $t=0$, the dipole moment is $0$. When relative error (MAPE) is calculated at this point, it causes numerical issues. The SMAPE is a more stable metric and time reversed SMAPE is similar to the baseline SMAPE.

\subsection{Runtime}

Inference is fast and consistent across encodings. Table~\ref{tab:input_representation} shows per-slice times around \SIrange{1.3}{1.5}{ms} on our test hardware for the ablations, and Tables~\ref{tab:coarse_comparison} and \ref{tab:generalizability} show timing around \SI{3}{ms} for baseline and generalization results. Even for spatial super-resolution with twice the number of spatial grid points, the time per step only increases to about \SI{3.37}{ms}. This millisecond-scale cost per step enables practical parameter sweeps and on-the-fly evaluation over varying laser fields.

\section{Discussion}
\label{sec:discussion}

The autoregressve FNO time propagator achieves accurate, stable density rollouts at millisecond per-step cost and clearly improves agreement with a high-resolution real time TDDFT reference compared to a coarse-grid numerical solver. In Table~\ref{tab:generalizability}, density MSE and dipole errors favor the FNO on the same coarse time grid, and Figure~\ref{fig:example_snapshot} shows that errors do not visibly accumulate beyond the input window. These gains are practical for rapid parameter sweeps where many trajectories must be evaluated.

Input design for the FNO model matters. Table~\ref{tab:input_representation} shows that encoding laser history alongside density substantially reduces both field-level errors and dipole errors relative to density-only baselines, without increasing inference time. The fully connected encodings perform best among the tested options, suggesting that separating density and laser features before combination is beneficial.

The model captures key observables with good fidelity. The dipole tracks the reference oscillations during the pulse (Figure~\ref{fig:dipole_energy}, left). Total energy trends under an approximate Thomas-Fermi proxy are reproduced qualitatively (Figure~\ref{fig:dipole_energy}, right) indicating that errors do not accumulate with simulation time. Because percentage errors can be inflated near dipole zero crossings, we report SMAPE alongside MAPE in Tables~\ref{tab:input_representation} and \ref{tab:generalizability}. Particle number remains close to the target value across rollouts due to the integral loss (Figure~\ref{fig:integral_error}). Empirical backward rollouts with reversed fields retrace the reference trajectory closely over the tested window (Figures~\ref{fig:tr_snapshot},~\ref{fig:tr_dipole_energy}), indicating approximate satisfaction of time-reversal symmetry of the FNO model in practice.

The FNO approach has clear limits. The demonstration is restricted to one-dimensional, two-electron diatomic molecules under ALDA and sinusoidal driving in the dipole approximation. However, our approach does generalize methodologically to three dimensions, but its feasibility needs yet to be shown. The operator advances densities directly and is not unitary at the orbital level, so constraints beyond an integral penalty are desirable. Data quality and boundary handling in the references can influence learning, even though trajectories with artefacts were filtered. While the TF approximation was used to validate energy conservation in an orbital-free manner, using orbital-dependent Kohn-Sham total energy formulation would be required for more accurate energy calculations. Extending FNO-based propagators to three-dimensional TDDFT is conceptually straightforward but computationally demanding. The FFT-based operator scales as $\mathcal{O}(n \log n)$ but the memory consumption of 3D feature tensors grows rapidly and is the primary bottleneck. Recent variants introduce architectural changes to mitigate this, enabling applications to a range of three-dimensional simulation tasks \cite{li_geometry-informed_2023,rahman_sparsified_2024,wang_prediction_2024}.

These results are most useful in two regimes. First, amortized acceleration: once trained, the propagator supports fast parameter scans and on-the-fly exploration. Second, hybrid solvers: a predictor–corrector strategy where the FNO proposes large steps and a conservative integrator periodically corrects simulation drift can reduce total step counts while preserving strict physical checks.

Future work should enforce additional physics (enforcing continuity, non-negativity, time-reversal symmetry loss), broaden coverage to richer laser pulse families and go beyond ALDA, and scale to higher dimensions and many-electron systems. Using the learned map for adaptive step sizing or as a preconditioner inside implicit schemes is a promising path to combine speed with rigorous numerical guarantees (i.e., stability, convergence and error bounds).

\section{Conclusion}
\label{sec:conclusion}

We show that machine learned time propagators have the potential to accelerate TDDFT calculations for one-dimensional diatomic molecules. The propagator generalizes well to longer time roll-out periods and higher resolution spatial grids while maintaining physical consistency and observable accuracy. Extending this work to three-dimensions would enable on-the-fly modeling of the electronic response properties of laser-excited molecules and materials in various scattering experiments that are conducted at photon sources around the globe. This would enable fast simulations that generalize well over the input parameters of the experimental setup. Rapid modeling would also enable the design of laser pulses to precisely control quantum dynamics under quantum optimal control theory \cite{werschnik_quantum_2007}.

\section*{Acknowledgments}
This work was supported by the Center for Advanced Systems Understanding (CASUS), which is financed by Germany’s Federal Ministry of Research, Technology and Space (BMFTR) and by the Saxon State government out of the State budget approved by the Saxon State Parliament. We acknowledge funding from the Helmholtz Association’s Initiative and Networking Fund through Helmholtz AI. 

\section*{Data Availability Statement} 
The data that support the findings of this study are available at \href{https://rodare.hzdr.de/record/3995}{RODARE record 3995, url: https://rodare.hzdr.de/record/3995}.

\section*{Conflict of Interest} 
The authors declare that they have no known competing financial interests or personal relationships that could have appeared to influence the work reported in this paper.

\section*{References} 
\bibliography{mlst_tddft_fno}
\bibliographystyle{unsrt}

\section*{Supplementary Information}
\label{sec:SI}

\subsection{Data Generation}
We model a one-dimensional diatomic (double-well) external potential
\begin{equation}
    v_{\text{ion}}(\mathbf{r}) = v_{\text{ion}}(x) = -\frac{Z_1}{\sqrt{(x-\frac{d}{2})^2+a^2}} - \frac{Z_2}{\sqrt{(x+\frac{d}{2})^2+a^2}},
\end{equation}

and include a time-dependent laser under the dipole approximation,
\begin{equation}
  v_{\mathrm{las}}(x,t) = A \sin(\omega t).
\end{equation}

The training domain and laser settings are summarized in Table~\ref{tab:data-specs}.

\begin{table}[h!]
\centering
\caption{Dataset specifications.}
\label{tab:data-specs}
\begin{tabular}{@{}ll@{}}
\toprule
Spatial domain & radius \([\!-9, 9]\) au,\quad grid spacing $\Delta x = \SI{0.05}{au}$ \\
Temporal domain & $t \in [0, \SI{5}{\femto\second}]$ \\
ML time step & $\Delta t_{\mathrm{ML}} = \SI{0.1}{\femto\second}$ \\
Reference time step & $\Delta t_{\mathrm{ref}} = \SI{0.01}{\femto\second}$ \\
Laser wavelength & \SIrange{400}{750}{\nano\meter} (optical range) \\
Laser intensity & \SI{e12}-\SI{e14}{W\per\centi\meter\squared}~\cite{andTDDFTStudyExcitation2013,bubinElectronionDynamicsLaserassisted2011a,wozniakChapterThreeExploring2023}\\
\bottomrule
\end{tabular}
\end{table}

\subsection{Tensor Shapes and Sequence Lengths}
We denote the number of spatial points by $S$, the number of autoregressive output steps by $T$, and the number of conditioning input steps by $T_{\mathrm{in}}$.

\begin{table}[h!]
\centering
\caption{Grid and sequence settings.}
\label{tab:shapes}
\begin{tabular}{@{}ll@{}}
\toprule
Spatial points & $S = 361$ \\
Output steps & $T = 41$ \\
Input steps & $T_{\mathrm{in}} = 10$ \\
Total horizon & $T_{\mathrm{total}} = 51$ \\
Grid spacing & $\Delta x = \SI{0.05}{au}$ \\
ML time step & $\Delta t = \SI{0.1}{\femto\second}$ \\
\bottomrule
\end{tabular}
\end{table}

\subsection{Training Parameters}
Parameters of model training are summarized in Table~\ref{tab:train}.

\begin{table}[h!]
\centering
\caption{Training configuration.}
\label{tab:train}
\begin{tabular}{@{}ll@{}}
\toprule
Epochs & 800 \\
Batch size & 40 \\
Dataset split & Train/Val/Test = $800/150/200$ (from $2048$ total) \\
Optimizer & AdamW \\
Learning rate & $1\times 10^{-3}$ \\
Weight decay & $1\times 10^{-4}$ \\
Scheduler & CosineAnnealingLR ($\eta_{\mathrm{min}}=1\times 10^{-5}$) \\
\bottomrule
\end{tabular}
\end{table}

\subsection{Model: Fourier Neural Operator}
Model hyperparameters are summarized in Table~\ref{tab:model}.

\begin{table}[h!]
\centering
\caption{Model hyperparameters.}
\label{tab:model}
\begin{tabular}{@{}ll@{}}
\toprule
Width & 128 \\
Layers & 3 \\
Fourier modes & 32 \\
Input representation & separate fully connected add (density and laser branches, baseline) \\
Padding & 40 (finite boundary conditions) \\
Grid features & enabled \\
\bottomrule
\end{tabular}
\end{table}

\subsection{Data Transformations}
\begin{itemize}
  \item Density: Log transform, $\log_{10}n(x,t)$
  \item Density: Scaled to [-1,1]
\end{itemize}

\subsection{Losses}
\begin{itemize}
  \item Mean squared error (density)
  \item Integral loss (weight $=0.1$).
\end{itemize}

\subsection{Training Algorithm}
\begin{algorithm}[H]
    \caption{Autoregressive FNO Training}
    \begin{algorithmic}[1]
    \footnotesize
    \State \textbf{Inputs:} dataset $\mathcal{D}=\{\,(n(x,t),\,v_l(t))_{t=1}^{T_{\text{total}}}\,\}$ of trajectories
    \State \hspace{1.9em} operator $\mathcal{G}_\theta:\underbrace{\mathcal{X}\times\cdots\times\mathcal{X}}_{T_{\mathrm{in}}}\times\mathcal{V}\to\mathcal{X}$ (FNO with params $\theta$)
    \State \hspace{1.9em} horizon $T_{\text{out}}$, reference input $T_{\mathrm{in}}$, batch size $B$, weight $\lambda_{\text{int}}\ge 0$
    
    \For{epoch $=1,2,\dots$}
      \For{each mini-batch $\{ \text{traj}_i \}_{i=1}^B$}
        \State sample start indices $\{s_i\}_{i=1}^B$ with $T_{\mathrm{in}} \le s_i \le T_{\text{total}}-T_{\text{out}}$
        \State initialize window from reference data:
        \[
          \mathsf{win}_i \gets \big(n_i(x, s_i - T_{\mathrm{in}} + 1),\,\dots,\,n_i(x, s_i)\big)
        \]
        \State \textbf{Rollout:}
        \For{$\tau = 0,1,\dots,T_{\text{out}}-1$}
          \State form input $x_i^\tau \gets (\mathsf{win}_i,\; v_{l,i}(s_i+\tau))$
          \State predict next snapshot:
          \[
            \widehat{n}_i(x, s_i+\tau+1) \gets \mathcal{G}_\theta(x_i^\tau)
          \]
          \State update window (drop oldest, append prediction):
          \[
            \mathsf{win}_i \gets \big(\mathsf{win}_i[2{:}],\; \widehat{n}_i(x, s_i+\tau+1)\big)
          \]
        \EndFor
    
        \State \textbf{Loss over rollout:}
        \State $\displaystyle \mathcal{L}_{\mathrm{MSE}} \gets \frac{1}{B T_{\text{out}}}
          \sum_{i=1}^B \sum_{\tau=1}^{T_{\text{out}}}
          \| \widehat{n}_i(\cdot, s_i+\tau) - n_i(\cdot, s_i+\tau) \|_2^2$
        \State $\displaystyle \mathcal{L}_{\mathrm{int}} \gets \frac{1}{B T_{\text{out}}}
          \sum_{i=1}^B \sum_{\tau=1}^{T_{\text{out}}}
          \left( \int \widehat{n}_i(x, s_i+\tau)\,dx
                - \int n_i(x, s_i+\tau)\,dx \right)^2$
        \State total loss: \(\mathcal{L} \gets \mathcal{L}_{\mathrm{MSE}} + \lambda_{\mathrm{int}}\,\mathcal{L}_{\mathrm{int}}\)
    
        \State \textbf{Update:}  backpropagate $\nabla_\theta \mathcal{L}$
      \EndFor
    \EndFor
    \end{algorithmic}
    \end{algorithm}    
\subsection{Time Propagation Algorithm}
\begin{algorithm}[H]
    \caption{Autoregressive FNO Iteration}
    \begin{algorithmic}[1]
        \footnotesize
        % Initial condition / data
        \State \textbf{Given:} 
          \(\{n(x,1),\, n(x,2), \dots, n(x,T_{\mathrm{in}})\}\)
        \State \textbf{Given:} laser potentials 
          \(\{v_l(1), v_l(2), \dots, v_l(T)\}\)
        \State \textbf{Given:} 
          \(\mathcal{G}:\underbrace{\mathcal{X} \times \cdots \times \mathcal{X}}_{\smash{T_{\mathrm{in}}}} \times \mathcal{V} \to \mathcal{X}\) (FNO operator)
  
        % Main loop
        \For{\(t = T_{\mathrm{in}}, T_{\mathrm{in}}+1, \dots, T-1\)}
            \State \textbf{Form input:}
              \[
                \bigl( n(x, t - T_{\mathrm{in}} + 1), \dots, n(x, t), \, v_l(t) \bigr)
              \]
            \State \textbf{Predict next snapshot:}
              \[
                 \widehat{n}(x, t+1) \;=\;
                 \mathcal{G} \Bigl( n(x, t), \dots, n(x, t - T_{\mathrm{in}} + 1), v_l(t) \Bigr)
              \]
            \State \textbf{Update state:}
              \[
                 n(x, t+1) \;\leftarrow\; \widehat{n}(x, t+1)
              \]
        \EndFor
    \end{algorithmic}
    \end{algorithm}

\subsection{Compute environment}
All computations were performed on the HZDR Hemera HPC cluster, with Intex Xeon 2.4 GHz processors and Nvidia Tesla V100 32GB GPUs.

\end{document}